\documentclass[twocolumn,trackchanges]{aastex631}

\begin{document}

\title{Long-Term X-ray Variability on the Benchmark YSO HL Tau}


\author[0000-0002-3741-4181]{Steven M. Silverberg}
\affiliation{Smithsonian Astrophysical Observatory, MS 70, 60 Garden St., Cambridge, MA 02138}

\author[0000-0002-0826-9261]{Scott J. Wolk}
\affiliation{Smithsonian Astrophysical Observatory, MS 70, 60 Garden St., Cambridge, MA 02138}

\author[0000-0002-7939-377X]{David A. Principe}
\affiliation{MIT Kavli Institute for Astrophysics and Space Research, 77 Massachusetts Avenue, Cambridge, MA 02139, USA}

\author[0000-0002-5094-2245]{P. C. Schneider}
\affiliation{Hamburger Sternwarte, Gojenbergsweg 112, D-21029, Hamburg, Germany}

\author[0000-0003-4243-2840]{Hans Moritz G\"unther}
\affiliation{MIT Kavli Institute for Astrophysics and Space Research, 77 Massachusetts Avenue, Cambridge, MA 02139, USA}

\author[0000-0001-6072-9344]{Jinyoung Serena Kim}
\affiliation{Steward Observatory, University of Arizona, 933 N. Cherry Ave., Tucson, AZ 85721-0065, USA}

\author[0000-0002-3138-8250]{Joel H. Kastner}
\affiliation{Center for Imaging Science, School of Physics and Astronomy, and Laboratory for Multiwavelength Astrophysics, Rochester Institute of Technology, Rochester,
NY 14623, USA}

\begin{abstract}
HL Tau is one of the most well-studied Class I young stellar objects, including frequent observations at near- and mid-infrared, (sub-) millimeter, and X-ray wavelengths. We present the results of an X-ray variability monitoring campaign with \textit{XMM-Newton} in 2020 and X-ray gratings spectroscopy from \textit{Chandra}/HETGS in 2018. We find that the X-ray spectrum of HL Tau is consistently hot (with characteristic plasma temperatures $T \gtrsim 30$ MK) over 31 epochs spanning 20 years, which is consistent in temperature with most Class I YSOs. The high-resolution HETG spectrum indicates the presence of some cooler plasma. We characterize the variability of the star across the 31 observations and find a subset of observations with significant variability on a $\sim$21-day timescale in the observed count rate and flux. We discuss the possible origins of this variability, and identify further observations that would better constrain the nature of the changes.
\end{abstract}


\section{Introduction} \label{sec:intro}

Observations of pre-main-sequence stars and their circumstellar disks are essential for understanding the environment in which exoplanetary systems form and evolve. These systems initially form from the collapse of large molecular clouds into envelopes around a natal young stellar object with radial infall from the thick envelope (a Class 0/1 YSO). Over time, this envelope flattens to a circumstellar disk, revealing photospheric emission from the young pre-main sequence star (Class II; ``classical'' T Tauri star). Disk mass is eventually dissipated via planet formation, accretion of disk material onto the central star, and dispersal of material by photoevaporation driven by the central star. This process eventually results in a diskless T Tauri star (Class III). These young low-mass ($<2 M_{\odot}$) pre-main sequence stars exhibit a variety of phenomena that affect the evolution of their circumstellar material, including severe accretion events \citep[e.g.~FUOr and ExOr events][]{2014prpl.conf..387A}, jets, large molecular outflows, internal photoevaporation of the disk by high-energy photons from the central star, and planet formation \citep[e.g.][]{2016ARA&A..54..135H}.

One of the original defining characteristics of young stellar objects (YSOs) is their optical variability \citep{1945ApJ...102..168J,1952JRASC..46..222H}. A significant amount of this variability takes the form of periodic brightening and dimming, interpreted as hot and cool spots on the stellar surface rotating into and out of view \citep[e.g.~][]{1983ApJ...267..191R,1986ApJ...306..199V}. Since then, several focused surveys have been performed to understand the short- and long-term variability of these systems at a variety of wavelengths, leveraging the capability of space telescopes for surveys mostly unaffected by the diurnal cycle. The YSOVAR surveys \citep[][and references therein]{2014AJ....148...92R,2014AJ....148..122G,2015AJ....150..145W} focused on mid-infrared variability using \textit{Spitzer Space Telescope} observations, identifying several infrared-variable YSOs across eleven star-forming regions. The CSI 2264 program \citep{2014AJ....147...82C,2014AJ....147...83S} used \textit{CoRoT} and \textit{Spitzer} observations to identify a plethora of variability phenomena, including ``bursts,'' frequent aperiodic brightening events attributable to accretion events; ``dips,'' dimming events attributable to variability of the circumstellar disk obscuring the star; and the presence of a population of semi-regular variables similar to AA Tau \citep{1999A&A...349..619B,2015A&A...584A..51S,2015AJ....149..130S}.

The study of the Taurus star-forming region (Taurus SFR) fundamentally shapes our understanding of the processes of young star formation and evolution \citep[e.g.][]{1995ApJS..101..117K}. Its proximity to Earth \citep[$\sim 140$ pc; e.g.][and references therein]{2018ApJ...859...33G} makes even the lowest-mass stars and brown dwarfs available for examination. As such, Taurus is one of the most extensively studied star-forming regions, in X-rays \citep[XEST][]{2007AandA...468..353G}, mid-infrared \citep{2005ApJ...629..881H,2010ApJS..186..259R,2011ApJS..195....3F,2014ApJ...784..126E}, and sub-millimeter/millimeter \citep[e.g.][]{2018ApJ...869...17L,2020A&A...636A..65G}. Campaign 13 of the \textit{Kepler}-2 (K2) mission included the Taurus-Auriga complex, providing near-continuous white-light light curves of stars in the complex for nearly three months \citep{2022AJ....163..212C}.

Within the Taurus SFR, the Class I YSO HL Tau has become a benchmark target for studies of early disk and planet formation. HL Tau hosts one of the most thoroughly studied protoplanetary disks \citep[e.g.][]{1983ApJ...270L..69C,1984ApJ...283L..57G,1986ApJ...309..755B,1989A&A...215L...1M,1990AJ.....99..924B}. ALMA images of the system show concentric bright and dark rings, suggesting disk gaps \citep{2015ApJ...808L...3A}. There has been intense speculation as to the causes of the observed gaps, including potential sculpting by planets \citep[e.g.][]{2015ApJ...809...93D}, though non-planet explanations have also been proposed \citep[e.g.][]{2016A&A...590A..17R}. The system is one of the most well-studied YSOs at sub-mm and mm wavelengths  \citep[e.g.][]{2016ApJ...821L..16C,2024NatAs.tmp...49F}, but very little is known about the innermost portions of the disk, where dust is optically thick even at mm wavelengths \citep{2016ApJ...821L..16C}. HL Tau was observed with the Near-Infrared Camera (NIRCam) aboard the James Webb Space Telescope (JWST), but even these observations only provide information for the region beyond $0.2''$, or $\sim30$ AU \citep{2024AJ....167..183M}. Even estimates of the mass of the central star have been uncertain; modeling based on adaptive optics observations of the system have found a mass of $\sim0.7 M_{\odot}$ \citep{1997ApJ...478..766C}, while optical spectroscopy of the system has yielded a mass of $1.2 M_{\odot}$ \citep{2004ApJ...616..998W}.

Here we present analysis of recent X-ray observations of HL Tau with the X-ray Multimirror Mission (XMM-Newton) and Chandra X-ray Observatory, in conjunction with re-analysis of archival X-ray data. We find that the observed stellar X-ray spectrum is hot and heavily absorbed, and find indications of a $\sim$21-day period in the system's X-ray emission. 

In Section \ref{sec:observations} we summarize our observations, including re-analysis of archival X-ray observations of HL Tau. In  Section \ref{sec:results} we summarize our analysis of these new data, presenting best-fit plasma temperatures and considering X-ray variability over time. We discuss potential origins of the features we observe in Section \ref{sec:discussion}. Finally, we summarize our findings and present recommendations for future study of the system in Section \ref{sec:conclusion}.

\section{Observations}
\label{sec:observations}

Below, we briefly describe the new and archival observations and data reduction. We summarize our new X-ray observations in Table \ref{tab:obs_sum}.

\begin{deluxetable*}{lcccc}
\tablecaption{Summary of New Observations of HL Tau \label{tab:obs_sum}}
\tablehead{
\colhead{} & \colhead{} & \multicolumn{2}{c}{Start Date} & \colhead{Duration} \\
\colhead{obsid} & \colhead{Telescope/Instrument} & \colhead{(MJD)} & \colhead{(UTC)} & \colhead{(ks)}}
\startdata
21946      & \textit{Chandra}/ACIS/HETG & 58415.68640 & 2018-10-24 & 12.0 \\
20160      & \textit{Chandra}/ACIS/HETG & 58417.26343 & 2018-10-26 & 41.5 \\
21947      & \textit{Chandra}/ACIS/HETG & 58418.28612 & 2018-10-27 & 12.0 \\
21948      & \textit{Chandra}/ACIS/HETG & 58418.62041 & 2018-10-27 & 56.0 \\
20161      & \textit{Chandra}/ACIS/HETG & 58419.90608 & 2018-10-28 & 48.5 \\
21950      & \textit{Chandra}/ACIS/HETG & 58421.71834 & 2018-10-30 & 14.0 \\
21951      & \textit{Chandra}/ACIS/HETG & 58422.54435 & 2018-10-31 & 36.5 \\
21952      & \textit{Chandra}/ACIS/HETG & 58424.40864 & 2018-11-02 & 12.5 \\
21953      & \textit{Chandra}/ACIS/HETG & 58425.19608 & 2018-11-03 & 36.5 \\
21954      & \textit{Chandra}/ACIS/HETG & 58433.72045 & 2018-11-11 & 25.5 \\
21965      & \textit{Chandra}/ACIS/HETG & 58434.31770 & 2018-11-12 & 25.5 \\
0865040201 & \textit{XMM-Newton}/EPIC   & 59079.67552 & 2020-08-18 & 36.8 \\
0865040301 & \textit{XMM-Newton}/EPIC   & 59083.65674 & 2020-08-22 & 40.0 \\
0865040401 & \textit{XMM-Newton}/EPIC   & 59089.81463 & 2020-08-28 & 33.0 \\
0865040601 & \textit{XMM-Newton}/EPIC   & 59095.94792 & 2020-09-03 & 33.0 \\
0865040701 & \textit{XMM-Newton}/EPIC   & 59104.26799 & 2020-09-12 & 33.0 \\
0865040501 & \textit{XMM-Newton}/EPIC   & 59110.54296 & 2020-09-18 & 47.9 \\
\enddata
\end{deluxetable*}

\subsection{Previously Unpublished \textit{Chandra} Observations}

HL Tau was observed by \textit{Chandra}/ACIS eleven times over the span of three weeks, from 2018 October 24 through 2018 November 12, with the High-Energy Transmission Grating Spectrograph (HETGS)  \citep{2005PASP..117.1144C}. The aimpoint was centered between HL Tau and the nearby ($30"$ separation) XZ Tau AB, another X-ray active star \citep{2023AJ....166..148S}, with the goal of observing both sources in parallel. Data were reduced with the Chandra Interactive Analysis software (CIAO; v.\ 4.15.1). The observations were energy filtered (0.5-8.0 keV) and time-filtered on good time intervals to reduce flaring particle background. Zeroth-order and gratings spectra were extracted with standard procedures in CIAO.

\subsection{Previously Unpublished \textit{XMM-Newton} Observations}

HL Tau was observed by the X-ray Multimirror Mission (\textit{XMM-Newton}) observatory six times over the course of 33 days from 2020 August 18 through 2020 September 19, as part of a larger campaign to monitor the variability of young stellar objects in Taurus (PI Schneider). The observations used the medium-thickness optical blocking filter. XZ Tau and HL Tau were extracted using standard procedures in SAS version 20.0. The observations were energy-filtered (0.3-9.0 keV) and time-filtered on good time intervals to reduce flaring particle background.

Because of the close proximity of XZ Tau AB and HL Tau, we defined custom extraction regions to ensure minimal contamination of HL~Tau by XZ~Tau. XZ Tau AB strongly flared during two observations with \textit{XMM-Newton} \citep{2023AJ....166..148S}, such that the emission from XZ Tau overlaps HL Tau on the detector. To mitigate this, for all observations we selected as a background a partial annulus centered on XZ Tau with inner and outer radii corresponding to the diameter of the HL Tau extraction region, excluding a $50^{\circ}$ wedge centered on the extraction region to avoid self-subtraction of HL Tau, as shown in Figure \ref{fig:MOS2_image}. We estimate that this procedure leaves less than $2\%$ contamination.

\subsection{Archival Optical and X-ray Data}

To monitor the optical emission from the XZ Tau AB and HL Tau systems, we requested observations of both from the Association of Amateur Variable Star Observers (AAVSO) over the time periods of observation by \textit{Chandra} and \textit{XMM-Newton}. These observations, distributed across multiple observers at multiple locations, provide low-cadence optical monitoring over the course of the observations. We supplemented these data with $V$-band light curves from the All-Sky Automated Survey for Supernovae \citep[ASAS-SN,][]{2019MNRAS.486.1907J}, and from previous analysis of data on HL Tau from the K2 mission \citep{2022AJ....163..212C}. Examination of the HL Tau light curve from the \textit{Transiting Exoplanet Survey Satellite} (TESS) showed it to be dominated by light from XZ Tau AB, due to the relative brightness of the two targets and the $21"$ pixel scale of TESS relative to the $30"$ separation of the systems.

To build a longer baseline for our analysis and ensure a one-to-one correspondence between analysis of the new data and previous work, we also re-reduced observations of HL Tau from 2004 originally presented in \citet{2006A&A...453..241G}, and re-reduced observations of HL Tau from 2000 initially presented in \citet{2003A&A...403..187F} and re-analyzed in \citet{2007AandA...468..353G}. We reprocessed these data with SAS version 21.0.0 and its calibrations to better ensure compatibility between analyses, as past work has shown that different conclusions can be drawn from the same data set with different editions of the same reduction software, e.g.\ the different conclusion drawn on the 2000 \textit{XMM-Newton} observation of XZ Tau between \citet{2003A&A...403..187F} and \citet{2006A&A...453..241G}. Unlike the 2020 observations, \textit{HL Tau} strongly flared during the 2004 observations. We thus adopt a different region for background extraction for these data than we used for the 2020 data, as the 2020 background region would include photons from HL Tau and thus lead to self-subtraction.

\section{Results}
\label{sec:results}

\subsection{Spectral Characteristics}

\begin{figure}
    \centering
    \includegraphics[width=0.47\textwidth]{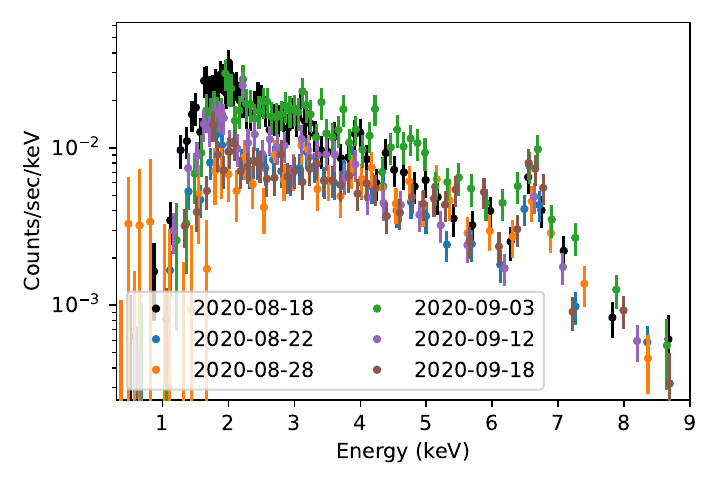}
    \caption{\textit{XMM-Newton} EPIC-PN spectra from each of the six $\sim40$-ks observations (represented by different colors as listed in the legend) in the the 2020 monitoring campaign. Spectra are binned by 15 counts per bin.}
    \label{fig:spectral_comparison}
\end{figure}

The new data confirm that, as expected from past X-ray observations and the Class I nature of the system, HL Tau is heavily absorbed in X-rays. As seen in Figure \ref{fig:spectral_comparison}, the pseudo-continuum of the spectrum varies over time at higher energies ($\gtrsim$ 1.5 keV) that are minimally affected by absorption, indicating variability intrinsic to the star, rather than changes in the absorption level. We adopt a model consisting of a single-temperature collisionally-excited plasma using the \texttt{xsvapec} \citep{2012ApJ...756..128F} as implemented in Sherpa with abundances from \citet{1989GeCoA..53..197A}, and a single absorption component, using the \texttt{xstbabs} absorption model \citep{2000ApJ...542..914W}. We link abundances for elements with a first ionization potential (FIP) similar to iron (i.e.\ Mg, Si, Ca, Ni) to the iron abundance, which we initially treat as a free variable along with plasma temperature, plasma normalization, and hydrogen column density. After this initial fit, we freeze the iron and iron-like-FIP abundance to the best-fit value for a joint fit to observations 0865030301, 0865040501, and 0865040701, the three faint \textit{XMM-Newton} observations that do not coincide with a flare from XZ Tau AB. Noting some apparent degeneracy between the variability pattern of temperature and $N_{H}$ in the 2020 observations, we also tested fitting with a fixed single temperature and varying hydrogen column and normalization, under the assumption that any change in temperature was instead due to a change in $N_{H}$. However, our best fits to the archival observations from 2000 and 2004 indicate that in some cases the temperature is lower without a corresponding increase in $N_{H}$, and so we adopt the variable-single-temperature option.

\begin{figure}
    \centering
    \includegraphics[width=0.4135\textwidth]{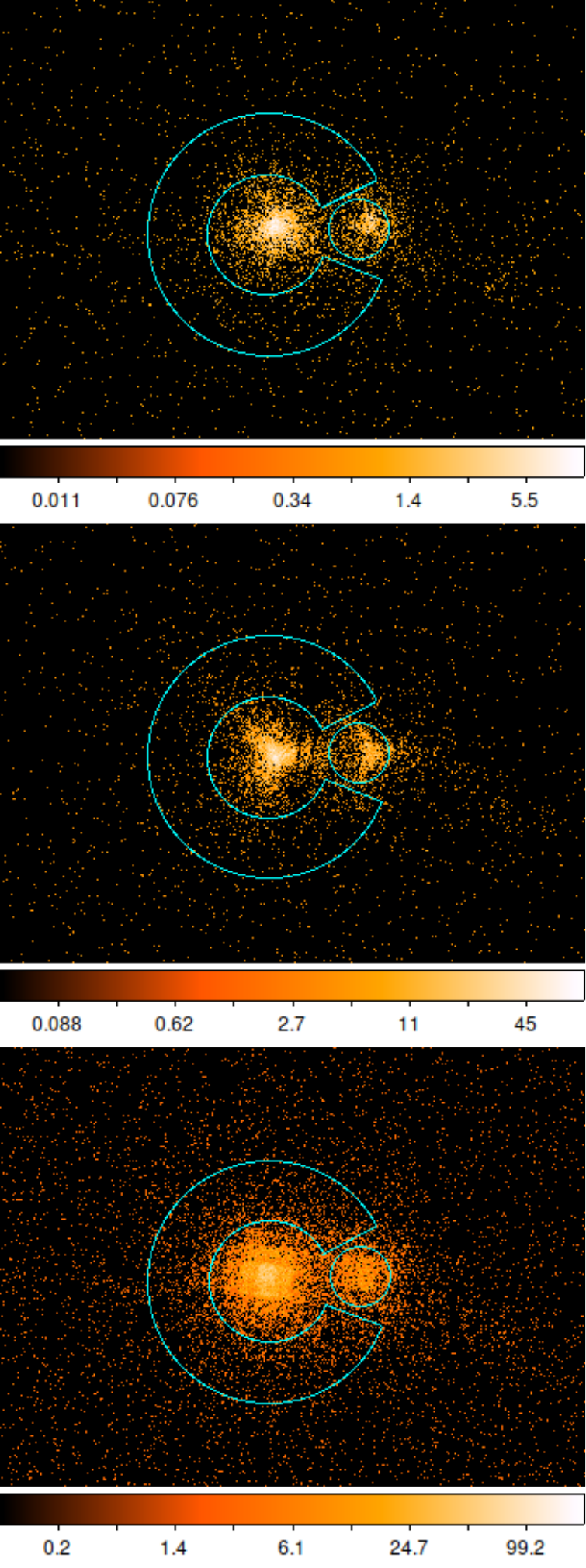}
    \caption{X-ray images of HL Tau from \textit{XMM-Newton} EPIC observation 0109060301. The MOS2 image (center) shows contamination of the core of the HL Tau (right source) from the wing of the PSF of XZ Tau AB (left source). This does not occur for the MOS1 (top) or PN (bottom) detectors. Cyan contours depict the extraction region for HL Tau (circle) and the background (partial annulus around XZ Tau).}
    \label{fig:MOS2_image}
\end{figure}

For the \textit{XMM-Newton} data, we simultaneously fit spectra from the EPIC PN, MOS1, and MOS2 detectors, with the exception of observations 0865040401 and 0109060301. In these two observations (as shown in Figure \ref{fig:MOS2_image}), the orientation of the telescope is such that the triangular PSF of XZ Tau points directly into the core of the HL Tau PSF for the MOS 2 detector, thoroughly contaminating it. As such, we only fit the MOS1 and PN spectra during these observations. For each individual \textit{Chandra}/HETG-S observation, we simultaneously fit the zeroth-order spectrum, the combined first-order MEG spectrum, and the combined first-order HEG spectrum.

\begin{deluxetable*}{lccccccccc}
\tablecaption{Fit Results with Fixed Iron and Iron-Like Abundances\tablenotemark{a}}
\label{tab:modelresults}
\tablehead{\colhead{} & \colhead{} & \colhead{CentralTime} & \colhead{Duration} & \colhead{$N_{H}$} & \colhead{kT} & \colhead{EM} & \multicolumn{2}{c}{Flux ($10^{-13} \mathrm{erg\,s^{-1}\,cm^{-2}}$)} & \colhead{log($L_{X}$)\tablenotemark{c}} \\
\colhead{ObsID} & \colhead{rstat\tablenotemark{b}} & \colhead{(MJD)} & \colhead{(days)} & \colhead{($10^{22} \mathrm{cm}^{2}$)} & \colhead{(keV)} & \colhead{($10^{52} \mathrm{cm}^{-3}$)} & \colhead{Absorbed} & \colhead{Unabsorbed} & \colhead{($\mathrm{erg\,s^{-1}}$)}}
\startdata
FaintXMM\tablenotemark{d}   & 0.449 &     \nodata & \nodata & $3.0_{ -0.1 }^{+ 0.2 }$ & $   3.9_{ -0.3 }^{+  0.3 }$ & $ 16_{  -1 }^{+  1 }$ & $ 4.2_{  -0.1 }^{+ 0.1 }$ & $10.2_{  -0.5 }^{+  0.4   }$ & $30.4$ \\
0865040201 & 0.435 & 59079.89628 & 0.40275 & $2.4_{ -0.1 }^{+ 0.1 }$ & $   2.9_{ -0.2 }^{+  0.2 }$ & $ 32_{  -2 }^{+  3 }$ & $ 7.1_{  -0.3 }^{+ 0.3 }$ & $18  _{  -1   }^{+  1     }$ & $30.6$ \\
0865040301 & 0.420 & 59083.89601 & 0.43978 & $3.4_{ -0.3 }^{+ 0.3 }$ & $   3.3_{ -0.3 }^{+  0.5 }$ & $ 18_{  -2 }^{+  3 }$ & $ 3.9_{  -1.0 }^{+ 0.9 }$ & $11  _{  -2   }^{+  2     }$ & $30.4$ \\
0865040401 & 0.463 & 59090.01340 & 0.35877 & $5.7_{ -0.9 }^{+ 1.1 }$ & $   3.5_{ -0.7 }^{+  1.2 }$ & $ 19_{  -5 }^{+  6 }$ & $ 3.5_{  -0.5 }^{+ 0.1 }$ & $11  _{  -2   }^{+  2     }$ & $30.4$ \\
0865040501 & 0.435 & 59110.82795 & 0.53120 & $3.2_{ -0.2 }^{+ 0.2 }$ & $   4.5_{ -0.5 }^{+  0.6 }$ & $ 15_{  -1 }^{+  2 }$ & $ 4.2_{  -0.7 }^{+ 0.7 }$ & $10  _{  -1   }^{+  1     }$ & $30.4$ \\
0865040601 & 0.477 & 59096.14666 & 0.35871 & $3.6_{ -0.2 }^{+ 0.3 }$ & $   3.0_{ -0.3 }^{+  0.3 }$ & $ 44_{  -5 }^{+  5 }$ & $ 8.5_{  -0.5 }^{+ 0.2 }$ & $25  _{  -2   }^{+  2     }$ & $30.8$ \\
0865040701           & 0.366 & 59104.46675 & 0.35876 & $2.4_{ -0.2 }^{+ 0.2 }$ & $   3.6_{ -0.4 }^{+  0.6 }$ & $ 18_{  -2 }^{+  2 }$ & $ 4.8_{  -0.5 }^{+ 0.1 }$ & $11  _{  -1   }^{+  1     }$ & $30.4$ \\
\textit{Chandra}\tablenotemark{e} & 0.298 &     \nodata & \nodata & $2.2_{ -0.2 }^{+ 0.2 }$ & $   2.7_{ -0.2 }^{+  0.3 }$ & $ 16_{  -2 }^{+  2 }$ & $ 3.4_{  -0.1 }^{+ 0.1 }$ & $ 8.7_{  -0.4 }^{+  0.4   }$ & $30.3$ \\ 
20160      & 0.433 & 58417.50142 & 0.47598 & $2.7_{ -0.5 }^{+ 0.6 }$ & $   3.0_{ -0.7 }^{+  1.3 }$ & $ 20_{  -5 }^{+  7 }$ & $ 5  _{  -2   }^{+ 1   }$ & $12  _{  -3   }^{+  3     }$ & $30.4$ \\
20161      & 0.361 & 58420.18519 & 0.55822 & $2.8_{ -0.5 }^{+ 0.5 }$ & $   3.3_{ -0.6 }^{+  1.0 }$ & $ 27_{  -6 }^{+  6 }$ & $ 6.2_{  -0.5 }^{+ 0.4 }$ & $16  _{  -1   }^{+  1     }$ & $30.6$ \\
21946      & 0.349 & 58415.75301 & 0.13323 & $3  _{ -2   }^{+ 2   }$ & $   2.2_{ -0.9 }^{+  4.3 }$ & $ 19_{ -12 }^{+ 32 }$ & $ 2.8_{  -2.4 }^{+ 0.1 }$ & $10  _{  -8   }^{+  8     }$ & $30.4$ \\
21947      & 0.211 & 58418.35311 & 0.13397 & $2.7_{ -0.9 }^{+ 1.2 }$ & $   2.2_{ -0.6 }^{+  1.3 }$ & $ 29_{ -12 }^{+ 21 }$ & $ 5  _{  -2   }^{+ 3   }$ & $15  _{  -6   }^{+  7     }$ & $30.5$ \\
21948      & 0.366 & 58418.94293 & 0.64506 & $1.7_{ -0.4 }^{+ 0.4 }$ & $   2.7_{ -0.5 }^{+  0.7 }$ & $ 13_{  -3 }^{+  3 }$ & $ 3.0_{  -0.6 }^{+ 0.1 }$ & $ 7  _{  -1   }^{+  1     }$ & $30.2$ \\
21950      & 0.358 & 58421.79902 & 0.16137 & $1.5_{ -0.7 }^{+ 1.2 }$ & $ > 3.6                   $ & $ 10_{  -2 }^{+  7 }$ & $ 4.6_{  -0.9 }^{+ 1.1 }$ & $ 7  _{  -1   }^{+  1     }$ & $30.2$ \\
21951      & 0.353 & 58422.75401 & 0.41933 & $2.2_{ -0.7 }^{+ 0.7 }$ & $   4.6_{ -1.3 }^{+  3.6 }$ & $ 12_{  -3 }^{+  4 }$ & $ 4  _{  -1   }^{+ 1   }$ & $ 8  _{  -2   }^{+  2     }$ & $30.2$ \\
21952      & 0.411 & 58424.47975 & 0.14222 & $2.7_{ -0.9 }^{+ 1.3 }$ & $   3  _{ -1   }^{+  3   }$ & $ 19_{  -7 }^{+ 14 }$ & $ 4.1_{  -2.2 }^{+ 0.3 }$ & $11  _{  -4   }^{+  3     }$ & $30.4$ \\
21953      & 0.423 & 58425.40648 & 0.42079 & $2.4_{ -0.5 }^{+ 0.6 }$ & $   2.8_{ -0.7 }^{+  1.0 }$ & $ 14_{  -4 }^{+  5 }$ & $ 3.1_{  -0.9 }^{+ 0.2 }$ & $ 8  _{  -2   }^{+  2     }$ & $30.3$ \\
21954      & 0.464 & 58433.86626 & 0.29160 & $0.9_{ -0.6 }^{+ 1.0 }$ & $   4  _{ -2   }^{+  7   }$ & $  5_{  -2 }^{+  4 }$ & $ 1.8_{  -1.8 }^{+ 0.1 }$ & $ 2.9_{  -2.7 }^{+  0.7   }$ & $29.8$ \\
21965      & 0.428 & 58434.46290 & 0.29040 & $1.5_{ -0.7 }^{+ 1.0 }$ & $   4  _{ -1   }^{+  4   }$ & $  7_{  -2 }^{+  4 }$ & $ 2.0_{  -1.5 }^{+ 0.1 }$ & $ 4  _{  -3   }^{+  1     }$ & $30.0$ \\
20906      & 0.393 & 58114.90951 & 0.45576 & $2.5_{ -0.3 }^{+ 0.2 }$ & $   2.4_{ -0.3 }^{+  0.5 }$ & $ 21_{  -3 }^{+  3 }$ & $ 3.8_{  -0.8 }^{+ 0.7 }$ & $11  _{  -2   }^{+  2     }$ & $30.4$ \\
18915      & 0.433 & 58124.74312 & 0.35868 & $2.3_{ -0.2 }^{+ 0.3 }$ & $   4.2_{ -0.7 }^{+  1.1 }$ & $ 27_{  -4 }^{+  4 }$ & $ 8.1_{  -0.5 }^{+ 0.3 }$ & $17  _{  -1   }^{+  1     }$ & $30.6$ \\
0200810201 & 0.332 & 53068.72175 & 0.09406 & $2.4_{ -0.5 }^{+ 0.6 }$ & $   4  _{ -1   }^{+  2   }$ & $ 17_{  -4 }^{+  6 }$ & $ 5  _{  -2   }^{+ 2   }$ & $10  _{  -4   }^{+  4     }$ & $30.4$ \\
0200810301 & 0.425 & 53069.19595 & 0.10449 & $2.3_{ -0.5 }^{+ 0.6 }$ & $   3.0_{ -0.7 }^{+  1.0 }$ & $ 10_{  -3 }^{+  4 }$ & $ 2.4_{  -0.5 }^{+ 0.2 }$ & $ 6  _{  -1   }^{+  1     }$ & $30.1$ \\
0200810401 & 0.352 & 53069.75446 & 0.09407 & $2.8_{ -0.4 }^{+ 0.5 }$ & $   2.9_{ -0.6 }^{+  1.0 }$ & $ 17_{  -4 }^{+  5 }$ & $ 4  _{  -1   }^{+ 1   }$ & $10  _{  -2   }^{+  3     }$ & $30.4$ \\
0200810501 & 0.309 & 53070.09709 & 0.12567 & $2.4_{ -0.5 }^{+ 0.6 }$ & $   3.2_{ -0.8 }^{+  1.8 }$ & $ 14_{  -4 }^{+  5 }$ & $ 3.3_{  -2.0 }^{+ 0.1 }$ & $ 8  _{  -4   }^{+  2     }$ & $30.3$ \\
0200810601 & 0.350 & 53070.71620 & 0.09407 & $2.4_{ -0.5 }^{+ 0.6 }$ & $   2.8_{ -0.6 }^{+  0.9 }$ & $ 16_{  -4 }^{+  6 }$ & $ 3.5_{  -0.6 }^{+ 0.3 }$ & $ 9  _{  -2   }^{+  2     }$ & $30.3$ \\
0200810701 & 0.648 & 53071.14741 & 0.08130 & $2.5_{ -0.3 }^{+ 0.2 }$ & $  10  _{ -2   }^{+  6   }$ & $ 74_{  -8 }^{+  7 }$ & $29.7_{  -4.1 }^{+ 0.8 }$ & $54  _{  -7   }^{+  3     }$ & $31.1$ \\
0200810801 & 0.434 & 53071.66793 & 0.07222 & $2.9_{ -0.2 }^{+ 0.2 }$ & $   3.7_{ -0.5 }^{+  0.5 }$ & $152_{ -16 }^{+ 21 }$ & $38  _{ -10   }^{+ 8   }$ & $93  _{ -15   }^{+ 12     }$ & $31.3$ \\
0200810901 & 0.428 & 53072.15073 & 0.17509 & $3.2_{ -0.2 }^{+ 0.2 }$ & $   6.4_{ -0.8 }^{+  1.1 }$ & $ 97_{  -7 }^{+  8 }$ & $32.6_{  -1.9 }^{+ 0.7 }$ & $68  _{  -5   }^{+  4     }$ & $31.2$ \\
0200811001 & 0.431 & 53072.71759 & 0.09407 & $2.6_{ -0.4 }^{+ 0.4 }$ & $   2.8_{ -0.4 }^{+  0.6 }$ & $ 29_{  -6 }^{+  7 }$ & $ 5.9_{  -1.3 }^{+ 0.3 }$ & $16  _{  -4   }^{+  3     }$ & $30.6$ \\
0200811101 & 0.416 & 53073.12994 & 0.11373 & $2.3_{ -0.4 }^{+ 0.5 }$ & $   2.9_{ -0.6 }^{+  0.8 }$ & $ 17_{  -4 }^{+  5 }$ & $ 3.8_{  -0.8 }^{+ 0.2 }$ & $10  _{  -2   }^{+  2   }$ & $30.4$ \\
0200811201 & 0.482 & 53073.65782 & 0.09407 & $1.8_{ -0.7 }^{+ 0.6 }$ & $   5  _{ -2   }^{+ 18   }$ & $ 10_{  -3 }^{+  4 }$ & $ 3  _{  -2   }^{+ 2   }$ & $ 7  _{  -3   }^{+  3   }$ & $30.2$ \\
0109060301 & 0.410 & 51797.10522 & 0.64886 & $2.3_{ -0.2 }^{+ 0.2 }$ & $   2.7_{ -0.3 }^{+  0.3 }$ & $ 15_{  -1 }^{+  2 }$ & $ 3.1_{  -0.2 }^{+ 0.1 }$ & $ 8.1_{  -0.8 }^{+  0.7 }$ & $30.3$ \\
\enddata
\tablenotetext{a}{Abundances for Fe, Si, Mg, Ca, and Ni fixed at 0.48 of the solar elemental abundances as provided by \citet{1989GeCoA..53..197A}}
\tablenotetext{b}{Reduced $\chi^{2}$ based on the Gehrels-weighted $\chi^{2}$ \citep{1986ApJ...303..336G}}
\tablenotetext{c}{Unabsorbed luminosity derived from the unabsorbed flux and distance to the star (140 pc).}
\tablenotetext{d}{Joint fit of obsids 0865040301, 0865040701, and 0865040501.}
\tablenotetext{e}{Joint fit of all zeroth order \textit{Chandra}/HETGS observations and the combined first order MEG and HEG spectra taken simultaneously.}
\end{deluxetable*}

We list the results of our fitting in Table \ref{tab:modelresults}. These consistently show that the X-ray emission from HL Tau is generally hot, generally bright, and quite variable in flux.

\subsection{X-ray Time Series}
\label{sec:timeseries}

\begin{figure*}
    \centering
    \includegraphics[width=0.98\textwidth]{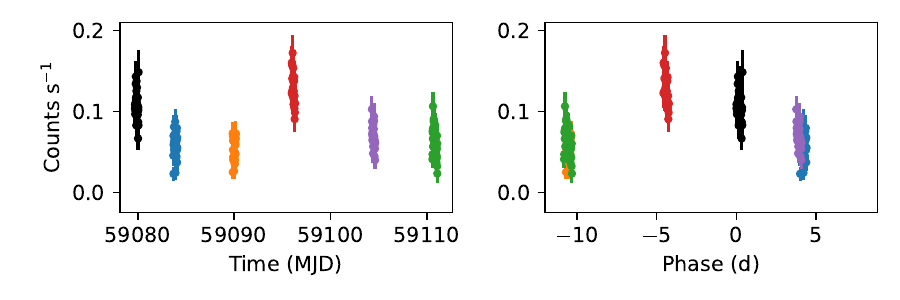}
    \caption{The high-energy ($>2$ keV) XMM/PN light curve of HL Tau from 2020, in chronological order (left) and phased to a $\sim$21-day period (right). Colors distinguish data from each observation and are the same for each observation in both plots. Count rates are derived from binning by 1 ks.}
    \label{fig:hltau_lightcurve_period}
\end{figure*}

\begin{figure}
    \centering
    \includegraphics[width=\linewidth]{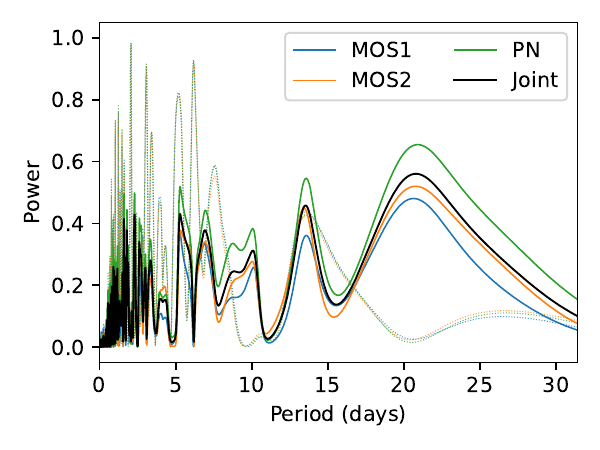}
    \caption{Lomb-Scargle periodograms of the high-energy light curves for HL Tau from the \textit{XMM-Newton} MOS1 (solid blue), MOS2 (solid orange), and PN (solid green) detectors and all three jointly (black). The Lomb-Scargle periodograms of the window functions (i.e.~flat data with identical timing and uncertainties to the observed data) for the three detectors are depicted as fainter dotted lines with the same colors as those derived from data. The strongest peak of the data is at a minimum of the periodogram of the window function, indicating that this is not due solely to the observing window. The strongest peak of the window function, at $\sim$2 days, corresponds to \textit{XMM-Newton}'s orbital period.}
    \label{fig:lomb_scargles_XMM}
\end{figure}

We analyzed both low- ($<2$ keV) and high-energy ($>2$ keV) light curves in count rate, for the \textit{XMM-Newton} and \textit{Chandra}/HETGS zeroth order data. Count rates were binned by 1 ks for \textit{XMM-Newton}, and 2 ks for \textit{Chandra} to account for \textit{Chandra's} comparatively smaller effective area. The epoch 2020 high-energy light curves for HL Tau from \textit{XMM-Newton}/EPIC show an apparent periodic fluctuation at a period of $\sim$21 days, as can be seen in in Figure \ref{fig:hltau_lightcurve_period}. To test this, we generated a generalized Lomb-Scargle periodogram (using the \texttt{fit\_mean} option in the \texttt{LombScargle} package in \texttt{astropy}) to the light curves from all three EPIC detectors, finding a peak in the power spectrum with a maximum corresponding to a period of 20.8 days. To confirm that this peak is not an artifact of the window function of our observations, we also generated the Lomb-Scargle periodogram of a constant flat signal with identical relative uncertainties to the observed data in the same observing windows---i.e.~the periodogram of the window function. The power spectra of the window function periodograms (Figure \ref{fig:lomb_scargles_XMM}) show that the peak we find in the periodogram of the observed data does not appear in the periodogram of the window function while other peaks in the periodogram of the observed data \textit{do} coincide with the peaks of the power spectrum of the window function. This indicates that the periodogram peak at $\sim$21 days is not merely an artifact of the observation pattern.

The statistical false-alarm probability measures the likelihood that a light curve of nothing but noise would produce a peak of comparable height to the peak of a given frequency. We estimate the false-alarm probability using the \texttt{astropy} implementation of the method of \citet{2008MNRAS.385.1279B}, and find that the false-alarm probability of a peak of this height at this frequency is on the order of $<10^{-27}$---that is, the probability that a data set that was \textit{only noise} would produce a peak of this magnitude is effectively nil \citep{2018ApJS..236...16V}. This does \textit{not} prove that the data set \textit{is} periodic at the strongest period given these observations, but indicates that there is, in fact, some underlying variability that is being detected.

Because of the breadth of the peak in the periodogram and the challenges of deriving an uncertainty on a period via the Lomb-Scargle method \citep[see discussion in][]{2018ApJS..236...16V}, we do not specify a Gaussian uncertainty on the period here, instead treating it as $\sim21$ days throughout. We expect based on experience that given our limited sample space the uncertainty in this period is on order $\sim0.2$ days.

\begin{figure}
    \centering
    \includegraphics[width=\linewidth]{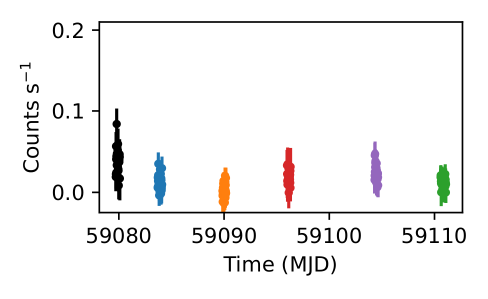}
    \caption{The low-energy ($<2$ keV) \textit{XMM-Newton} EPIC/PN light curve of HL Tau. Soft count rates are typically $<35\%$ of simultaneous hard count rates, due to the plasma temperature and high absorption.}
    \label{fig:lcXMMsoft}
\end{figure}

\begin{figure*}
    \includegraphics[width=\textwidth]{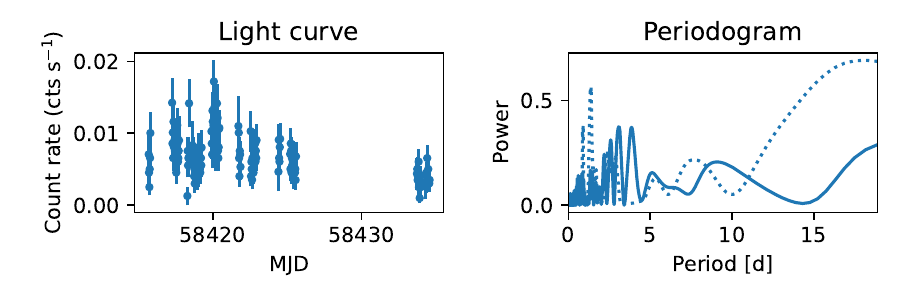}
    \caption{\textit{Left}: The zeroth-order light curve from the 2018 \textit{Chandra} observations. \textit{Right}: The Lomb-Scargle periodogram for this data (solid) and its window function (dotted). The strong peak in the window function near 2.5 days is consistent with the orbital period of \textit{Chandra}.}
    \label{fig:lcChandra}
\end{figure*}

By contrast, the low-energy light curve for \textit{XMM-Newton} (shown in Figure \ref{fig:lcXMMsoft}) shows low emission levels with no clear pattern to the fluctuations, consistent with the heavy absorption apparent in the system.\textit{Chandra} zeroth-order data (shown in Figure \ref{fig:lcChandra}) do not show a clear signature of periodicity on any timescale, though a periodogram of these data is not sensitive to periods longer than the observational baseline of $\sim19$ days. None of the individual observations in either the 2018 \textit{Chandra} campaign or the 2020 \textit{XMM-Newton} campaign appear to exhibit the typical fast rise and gradual decay of a flare, as was seen across obsids 0200810701-0200810901 in the 2004 observations of HL Tau \citep{2006A&A...453..241G}.

\subsection{Gratings data}

\begin{figure*}
    \includegraphics[width=\textwidth]{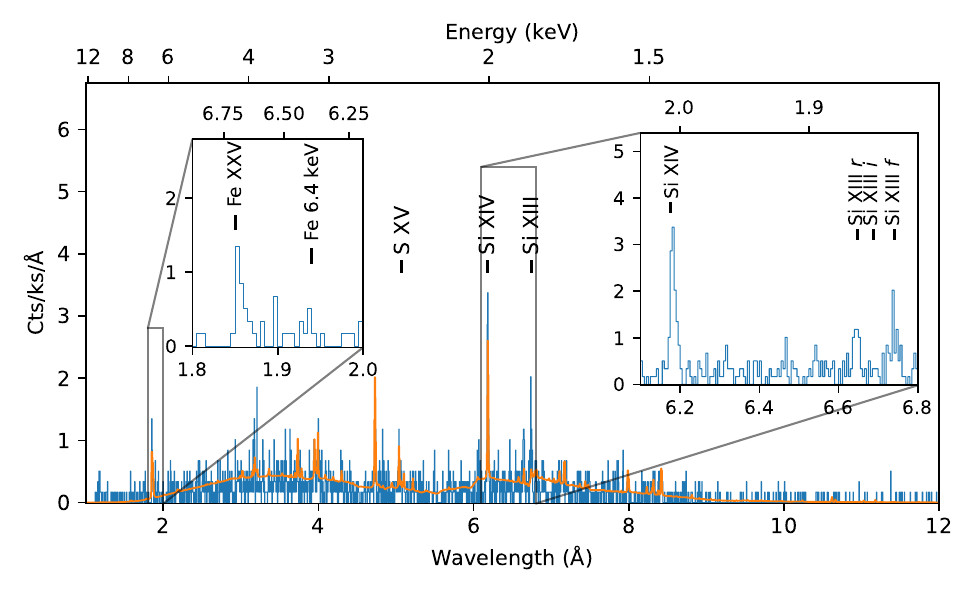}
    \caption{Combined gratings data from all 11 \textit{Chandra}/ACIS/HETG observations of HL Tau, with spectral lines of interest labeled. The best-fit model fit jointly to the individual zeroth order and combined gratings spectra is shown in orange. \textit{Left inset}: a zoom-in on the region around the Fe XXV feature at 1.85 \AA~(6.7 keV) shows a clear detection of the Fe XXV line and a marginal detection of the 6.4 keV neutral Fe fluorescence line in $\sim$300 ks. \textit{Right inset}: a zoom-in on the 6-7 \AA\ region shows a strong detection of the $f$ line of the Si XIII He-like triplet, and a fainter but still present detection of the $r$ line. Inset units are same as the larger figure.}
    \label{fig:gratings_fig}
\end{figure*}

We jointly fit combined (plus- and minus-) first order spectra from MEG and HEG with the zeroth-order CCD-resolution spectrum for each individual observation with \textit{Chandra}/HETG, to produce the results presented in Table \ref{tab:modelresults}. We assumed abundances identical to the best-fit abundances found in the \textit{XMM-Newton} data, as the signal-to-noise in the \textit{Chandra} data were insufficient for abundance determination. The combined gratings spectrum for each individual observation does not have sufficient signal-to-noise to effectively fit lines present in those data, so we combined \textit{all} the gratings data into one spectrum presented in Figure \ref{fig:gratings_fig}. While this approach loses time resolution, it allows for clearer identification of specific spectral lines, including strong emission from Si XIV and SI XIII, and a clear Fe XXV line.

\section{Discussion}
\label{sec:discussion}

\subsection{A particularly hot, bright CCD-resolution spectrum}

The first striking feature of HL Tau is the particularly hot spectrum. There is some degeneracy between the circumstellar absorption and the plasma temperature in the global model fitting (as the absorption could mask softer emission from the model). However, each of the \textit{XMM-Newton} spectra exhibit at least some emission from the Fe complex in the 6.4-6.7 keV range. Fe emission at 6.7 keV is \textit{only} produced with a significant amount of hot plasma---by contrast, spectra from the neighboring XZ Tau AB only exhibit this emission during significant flare events \citep{2023AJ....166..148S}. This independently confirms that there is a significant and relatively steady amount of plasma in HL Tau at temperatures higher than $>1$ keV.

While young stellar objects typically possess plasma with lower temperatures ($\sim$0.8 keV), temperatures in this range are not unheard of---they have been detected in the Orion Nebula Cluster \citep[ONC;][]{2005ApJS..160..401P}, and others. Such temperatures tend to be detected in less-evolved systems, but this is not seen consistently across all star-forming regions. While there is some degeneracy in determining whether the plasma is intrinsically hot or simply appears to be hot due to absorption of cooler plasma by the surrounding envelope, self-consistent methods at least mitigate heterogeneity within the analysis, enabling a first-order comparison. For instance, in $\rho$ Oph, \citet{2010A&A...519A..34P} found that the median temperature for the nine detected Class I YSOs in their sample was 4.4 keV. By contrast, \citet{2008ApJ...677..401P} found that the median of the average plasma temperature of Class 0-1a and Class 0-1b samples in Orion from the Chandra Orion Ultradeep Project (COUP) were $\sim2-3$ keV, and showed no trend in typical plasma temperature as a function of YSO class.

In the 2007 \textit{XMM-Newton} Extended Survey of Taurus (XEST) survey of young stars, HL Tau was found to have an average plasma temperature of 2.3 keV, the median of all Class I YSOs in that survey, based on an observation taken in 2000 and first presented in \citet{2003A&A...403..187F}. The sample from XEST also indicates that Class I YSOs tend to have hotter plasmas than their older class II counterparts. However, to achieve an acceptable fit, \citet{2007AandA...468..353G} modified their standard abundances for Taurus to an iron abundance of 1.06$\times$ the solar abundance. By contrast, our analysis of the 2000 \textit{XMM-Newton} observations that were used in XEST, with a fixed Fe abundance of $0.48 \pm 0.08$ of the solar value based on the faint 2020 observations, shows a temperature of $2.7 \pm 0.3$ keV, with an absorption of $2.3 (\pm 0.2) \times 10^{22} \mathrm{cm}^{2}$. We believe that this discrepancy is due to the use in \citet{2007AandA...468..353G} of the Wisconsin cross-sections \citep{1983ApJ...270..119M} rather than more modern absorption models, and the recalibration of the \textit{XMM-Newton} effective areas since the publication of XEST. 

\begin{figure}
    \includegraphics[width=0.473\textwidth]{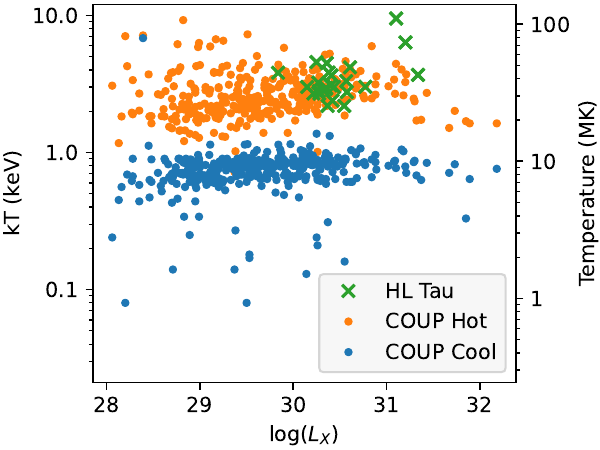}
    \caption{Temperature vs.\ unabsorbed X-ray luminosity for single-temperature fits to HL Tau (green 'X' markers), in comparison to temperature and unabsorbed luminosities from two-temperature fits (blue and orange points) to the COUP pre-MS stars \citep{2005ApJS..160..401P}.}
    \label{fig:COUP_comparison}
\end{figure}

To place our single-temperature evaluation of the data from HL Tau in further context, we also present our single X-ray plasma temperatures and their corresponding luminosities against the two-temperature models as a function of X-ray luminosity presented in the COUP \citep{2005ApJS..160..401P}, in Figure \ref{fig:COUP_comparison}. We see here that our single-temperature X-ray models are all consistent with the ``hot'' components of those stars observed in COUP. Our single-temperature model does not consider the impact of a cooler X-ray component, as such a component would be thoroughly absorbed; such a component would push the unabsorbed luminosities for HL Tau further to the right in this plot. Indeed, fitting the HL Tau data with a two-component model with one temperature fixed at 0.76 keV \citep[the steady-state cool component found in XZ Tau;][]{2023AJ....166..148S} increases the unabsorbed luminosity of the best-fit model to the faint \textit{XMM-Newton} observations by $\sim$0.1 dex. From this we assess that HL Tau remains a particularly bright X-ray source, albeit not an extreme case. The variability we observe further bolsters the hypothesis from \citet{2023AJ....166..148S} that single-snapshot surveys of young clusters are more indicative of the variability space that individual sources can fill, rather than a fixed indication of the steady state of individual T Tauri stars.

\subsection{Evidence of Cooler Plasma and Iron Fluorescence in High-Resolution Spectroscopy} 

We identify several apparent emission lines in the combined \textit{Chandra}/HETGS spectrum. The most notable among these is the Fe XXV line at 1.85 \AA---i.e.~the 6.7 keV line. This line appears in the combined spectrum and the vast majority of the individual observations, indicating that there is hot plasma present at all times through these observations, despite no evidence of flare activity.

We also detect emission lines from S XV, Si XIV, and Si XIII. However, the transitions we detect of these are \textit{not} always the brightest lines related to those populations. Notably, as shown in the right inset in Figure \ref{fig:gratings_fig}, while we do see some emission from the Si XIII recombination transition at 6.65 \AA, the brighter emission comes from the forbidden transition at 6.74 \AA, despite the $\mathbf{f}$ line having only slightly more than half the emissivity of the $\mathbf{r}$ line. The excitation rate for the $\mathbf{f}$ line peaks at a lower temperature than the rate for the $\mathbf{r}$ line. Overall, this indicates that the Si XIII emission comes from cooler plasma. Indeed, jointly fitting the two merged gratings spectra (from MEG and HEG) without zeroth-order data yields a best-fit single-temperature plasma of $3.1 \pm 0.2$ keV, compared to a best-fit single-temperature plasma for a joint fit of the zeroth-order spectra taken simultaneously to the gratings spectra of $3.9_{-0.4}^{+0.3}$ keV, a temperature difference of $0.8 \pm 0.4$ keV for simultaneous observations of the same system. The best fit to all of the \textit{Chandra}/HETGS data---both the individual zeroth order fits and the combined gratings spectra---is shown in Figure \ref{fig:gratings_fig}.

While we clearly detect the Fe XXV line at 6.7 keV, produced by hot plasma, in the gratings data, we also see a marginal detection of the 6.4 keV Fe K$\alpha$ line (Figure \ref{fig:gratings_fig}, left inset). A global fit to the spectrum including a Gaussian emission component at 6.4 keV would trivially produce a fit to the data with lower $\chi^{2}$ due to the lower number of degrees of freedom. Instead, we evaluate the detection of the source following the methodology outlined in \citet{2010ApJ...719..900K}, which identifies the minimum counts for detection of a source based on the background, using spectral binning in one dimension for the ``source'' (line) and ``background'' (surrounding continuum). We find seven counts in the line (between 1.925 and 1.95 \AA), which meets the minimum threshold for detection at a 10\% chance of false detection (false positive) and a 50\% chance of missed detection (false negative), based on the observed ``background'' continuum on either side of the line.

The Fe line at 6.4 keV has been found to occur in systems with hard X-ray emission and circumstellar disks \citep{2005ApJS..160..503T}, where emission from the 6.7 keV Fe complex and hotter causes cold iron in the disk to fluoresce. \citet{2005ApJS..160..503T} hypothesized that detection of the fluorescent line is constrained by the inclination angle of the system. However, \citet{2019A&A...623A..67P} noted that the number of detections of this line across different YSOs throws this into question, as one would not expect \textit{all} of those detections to be conveniently oriented for transmission of the line. Regardless, HL Tau's orientation \citep[$i \sim 47^{\circ}$;][]{2015ApJ...808L...3A} is favorable to detection of this line. Fluorescent line emission directly correlates with coronal emission in other Class I YSOs \citep[e.g.][]{2019A&A...623A..67P}.

The 6.4 keV Fe fluorescence line could be invaluable for diagnosing the origin of the variability; while \citet{2010A&A...520A..38C} found that K$\alpha$ emission was a quasi-persistent feature of some YSOs, including at quiescence, K$\alpha$ emission has been shown to directly correlate with coronal emission in Class I YSOs \citep[e.g.~][]{2019A&A...623A..67P}. Variations in the flux of the 6.4 keV line correlated with changes in the 6.7 keV flux would indicate variation intrinsic to the star, while stable 6.4 keV emission with varying observed 6.7 keV emission would indicate that the 6.7 keV emission originates from a feature on the stellar surface rotating into and out of view (e.g.~a stable hotspot), while continually illuminating the disk. Unfortunately, the line is too faint in our data for a time-resolved analysis. We do not clearly detect the 6.4 keV line in the spectra from \textit{XMM-Newton}.

\begin{figure*}
    \includegraphics[width=\textwidth]{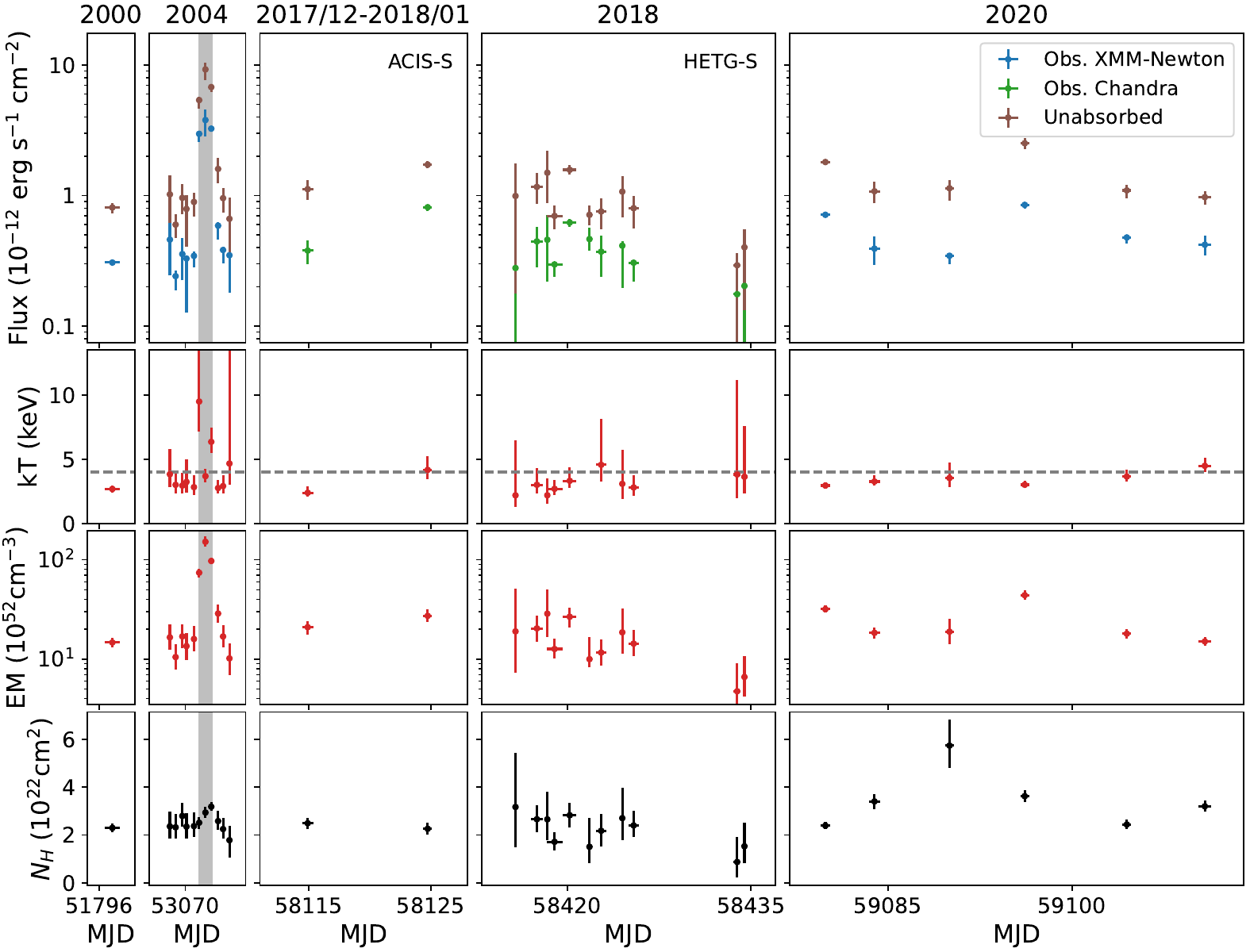}
    \caption{Comparison of model fits to HL Tau observations from 2000 to 2020. All models use one APEC plasma component with variable temperature and a fixed abundance for Fe (and elements with Fe-like first ionization potential) set at the best fit value from a joint fit to the faint XMM-Newton observations from 2020, and a variable absorption component. \textit{Top row}: Flux is shown both absorbed (as observed; blue for \textit{XMM-Newton}, green for \textit{Chandra}) and unabsorbed (corrected for $N_\mathrm{H}$; brown).
    \textit{Second row}: The plasma temperature for the single plasma component (red) in comparison to 4 keV. \textit{Third row}: the emission measure for the single plasma component (red).
    \textit{Bottom row}: the column density $N_{H}$ of the absorbing material (black). Observing windows are labeled above each column. The observations during the flare identified by \citet{2006A&A...453..241G} in 2004 are highlighted in gray.}
    \label{fig:modelsvtime}
\end{figure*}

\subsection{Characterizing the Variability}

\subsubsection{Variation in model components}

We examined temporal variability of the best fit model parameters for the data as a function of time, shown in Figure \ref{fig:modelsvtime}, to identify the underlying drivers of the change in emission. The best-fit models to the six epochs in 2020 clearly indicate that while there is change in the absorption column over time, consistent with evolution of the gas in the disk \citep[as discussed in][]{2024AJ....167..183M}, the change in the volume emission measure (the amount of emitting plasma) is more correlated with the change in the observed flux. Indeed, the best-fit plasma temperature in the 2020 data stays constant (within uncertainties) at $\sim$46 MK ($\sim$4 keV) across all observations. Rather than the changes in flux heralding a significant \textit{change} in the plasma temperature (e.g.~from a flare), the variability seems caused by changes in the \textit{amount} of plasma at the \textit{same} temperature.

\subsubsection{Context with archival observations}

To put the variability in the 2020 data set in context, we re-reduced and re-analyzed the archival observations from 2000 \citep{2003A&A...403..187F,2007AandA...468..353G}, 2004 \citep{2006A&A...453..241G}, and winter 2017-2018 \citep{2020ApJ...888...15S}, shown in Figure \ref{fig:modelsvtime} with the 2018 \textit{Chandra}/HETGS and 2020 \textit{XMM-Newton}/EPIC data. Due to the differences between detectors (the lower effective area of \textit{Chandra}/ACIS relative to \textit{XMM-Newton}/EPIC; the further diminishing of light-gathering power of the autumn 2018 \textit{Chandra} observations due to the presence of the gratings), we chose to look at the total observed flux from each observation. While this method loses sensitivity to variations on the order of kiloseconds by time-averaging across each full observation, it provides a consistent measure across observations and more robust statistical uncertainties on our measurements.

There are no clear signatures of periodicity at the $\sim$21 day period we identified in observations prior to 2020. While the observed fluxes from the winter 2017-18 data \textit{do} happen to fall in line with the periodic model extending back in time, the substantial uncertainty in the period suggests that this is as likely to be random coincidence as it is to be a true signal---a statistically-plausible difference in adopted period of 0.1 days leaves the period effectively unconstrained in any observations before 2020.

We similarly examined the best-fit model components for each observation across the twenty years of data and their change over time to identify any patterns of variation or consistency, akin to the stable cool-plasma component of the neighboring XZ Tau \citep{2023AJ....166..148S}). Unlike the case of XZ Tau, however, we find no clear consistency or pattern of change. The 2020 data from \textit{XMM-Newton} show clear variation in $N_{H}$, while the 2004 data show remarkable \textit{consistency} across five days of observation. While the 2020 data show hints of periodicity in the emission measure with consistent temperatures, the best-fit models to the 2017-18 \textit{Chandra} data indicate both variable emission measure and temperature.

Ultimately, each of these methods has their own form of bias, depending upon how we choose to model the system. While the data clearly exhibit patterns of variation, these are not always consistent with each other, nor do the components of our adopted models for each of these show patterns consistent with each other.

\subsubsection{Stochastic variability}

The simplest explanation for HL Tau's X-ray variability is stochastic---there is no pattern, but rather stochastic variability that happens to manifest as periodicity in our data. To test the likelihood of mere coincidence of finding a robust period in the data, we generated a series of ``fake'' light curves by drawing samples pseudo-randomly from the observed data. We generated light curves by randomly assigning (with replacement) the data from a real observation to a given block in time. We then randomly drew count-race/uncertainty pairs for each time in that block of time from the randomly-assigned obsid, with replacement. This allows randomization while preserving the differences in flux from observation to observation. We then evaluated the Lomb-Scargle periodogram for each of these randomly-generated light curves to see how commonly frequencies with stronger associated powers appeared in the data.

We found that such randomly-generated light curves produced higher peaks in the periodogram than the observed data more than 40\% of the time, suggesting that it might in fact be likely that the observed variation is simply by chance rather than anything periodic. However, this scenario leaves the question of what processes produce the changes, especially given the notable consistency of HL Tau's neighboring system.

One potential explanation would be stellar flares, triggered by a magnetic reconnection event of some form. While flares typically exhibit a fast-rise, exponential-decay (FRED) shape in light curves (e.g.~the 2004 flare on HL Tau), they can evolve more gradually, as seen in the ``slow-rise, top-flat'' flares observed in COUP \citep{2008ApJ...688..418G}.%

In the two bright observations from \textit{XMM-Newton} in 2020 (observations 0865040201 and 0865040601) the ``excess'' emission (emission above the unabsorbed flux from the ``faint'' observations) is consistent with luminosities of $\log(L_{X}) \sim 30.3$ and $\log(L_{X}) \sim 30.5$, {well below the excess luminosity of the 2004 HL Tau flare (peak excess luminosity $\log(L_{X}) \sim 31.3$). Both observations are consistent with a non-varying count rate (within ncertainties) at the 95\% confidence interval. However, many COUP ``slow-rise, top flat'' flares exhibited durations of $>100$ ks, well beyond the 40 ks we observe.}

\subsubsection{Exploring Non-stochastic Variability}

This data set cannot establish one way or the other whether the peak at $\sim$21 days in the Lomb-Scargle periodogram is indicative of an actual period on that timescale. However, a periodic signal on such a timescale would be of significant astrophysical interest were it to be identified on this star, or in another system in the future. We therefore find it worthwhile to explore the processes that \textit{could} give rise to recurring variability on a 21-day timescale.

Periodic signals from young stellar objects often indicate a feature rotating into and out of view, typically a hot or cold starspot, which are known to persist for multiple years on low-mass stars \citep[e.g.][]{2005MNRAS.357L...1B}. A $\sim$21-day sinusoidal period on HL Tau would indicate a feature near the equator \citep[system inclination angle $\sim$47$^{\circ}$, per][]{2015ApJ...808L...3A} moving into and out of view. Measuring the position and evolution of such cool and hot starspots and from them deriving a stellar rotation period has become common in the age of space-based time series photometry \citep[e.g.][]{2014AJ....147...82C, 2022AJ....163..212C}. However, stellar rotation periods (and thus rotation-driven starspot modulation) are typically less than this; \citet{2021ApJ...923..177S} found that of 1157 periodic variables in the ONC, none had periods longer than 14 days. Indeed, \citet{2004ApJ...616..998W} measured $v \sin i$ for HL Tau at $26.3 \pm 5.2 \mathrm{km\,s^{-1}}$. Assuming that the star rotates as a fixed body (i.e.~no differential rotation) and its rotation axis matches the inclination of the disk, the expected period of a surface feature at the equator would be $\sim$3.5 days. A 21-day timescale for variability indicates a significant deviation from the typical timescales of non-stochastic stellar variability.

Periods on the order of 10s of days, however, are more commonly seen as orbiting features around stars. \citet{2005ApJ...626..283B} followed the technique of \citet{2003ApJ...589..931N} to find an inner gas disk radius for HL Tau of 0.066 AU, based on the average velocity half-width at zero intensity (i.e.~the maximum velocity detected in the resolved line profile) of the $\nu = 2 - 1$ transition of CO emission and the assumption of Keplerian motion in a circular orbit, using an inclination angle of $71^{\circ} \pm 5^{\circ}$ \citep{2004MNRAS.352.1347L} and a stellar mass of 0.7 $M_{\odot}$ \citep{1997ApJ...478..766C}. Updating the inclination to $47^{\circ}$ \citep{2015ApJ...808L...3A} and the stellar mass to $1.2 M_{\odot}$ \citep{2004ApJ...616..998W} yields an inner gas disk radius of $\sim$0.07 AU. A $\sim$21-day period on HL Tau at its inclination angle would correspond to a circular Keplerian orbit at a radius of $\sim$0.15 AU, beyond the inner gas disk wall.

It is unclear what might orbit in the inner disk that could produce such a signature. Magneto-hydrodynamic (MHD) simulations of a star with a close-in planet show that interactions between a star and planet with its own magnetic field induce a hot spot on the star due to magnetic reconnection, at a specific point in the orbital phase of the planet \citep{2011ApJ...733...67C}; it is possible that an azimuthal asymmetry in the density of the inner disk could produce such activity, but that would likely require a unique configuration of the disk magnetic field to take the place of the planetary magnetic field. Alternatively, equatorial accretion might be an explanation. X-ray emission from accretion of disk material itself is typically cool \citep[from $\sim$0.3 keV plasma, e.g.][]{2002ApJ...567..434K, 2022hxga.book...57S}; the substantial absorption of HL Tau (Figure \ref{fig:spectral_comparison}) would hide emission at this low temperature from view. However, accretion-induced magnetic reconnection has been shown to produce hard-X-ray-bright hot spots on the outbursting protostar V1647 Ori \citep{2012ApJ...754...32H}. It is possible that what we see in the X-ray data of HL Tau is such a hot spot, driven by an accretion funnel from a fixed position in the inner disk.

MHD modeling of accretion disks around protostars have shown that disk currents emerge in systems where disk and stellar magnetic fields are anti-aligned \citep{2004A&A...420...17V}. This could produce the potential for continuously-occurring magnetic reconnection \textit{in the inner disk itself}. In addition to producing potential ongoing magnetic reconnection in the disk, such a system where the disk is anti-aligned to the stellar magnetic field emerges in MHD models of disk evolution where the initial seed magnetic field is that of a stellar dipole \citep{2015MNRAS.450..481D}, and such a system is expected to suppress disk winds due to inefficient mass loading. However, an inner disk radius of $\sim$0.07 AU would tend to preclude a feature moving into and out of view. The MIST evolutionary tracks \citep{2016ApJS..222....8D,2016ApJ...823..102C} indicate that a $1.6 M_{\odot}$ star at age $\sim1$ Myr will have a radius of $\sim 2.5 R_{\odot}$, such that any feature in the inner disk would be continuously visible at an inclination angle of $47^{\circ}$---it would not move behind the star.

\subsubsection{Further observations}

While the 2020 data contain tantalizing hints of recurrent variability, it is not possible to find more in the data than hints. The limited nature of the observations that signaled the potential for periodicity (six observations across 32 days) and the intrinsic uncertainty in frequency (due in part to time-averaging in each observation) means that a frequency cannot be tightly constrained with our data. Additional observations over a longer baseline would be capable of better evaluating characteristics of the variability.

Analysis of the system would also benefit from additional high-resolution (i.e.~gratings or microcalorimeter) X-ray spectroscopy of the system. We do not capture the cool plasma identified in the 2018 \textit{Chandra}/HETGS spectrum of the system with global models of CCD-resolution spectroscopy---repeated high-resolution spectroscopy would allow characterization of the evolution of this material over time in relation to the global variability in emission. The Fe fluorescence line at 6.4 keV could be invaluable for such observations---as this line emits from cold iron in the disk, rather than the stellar plasma, a comparison of the variability of that line relative to the variability of the observed stellar plasma spectrum could indicate whether the observed stellar variability is due to changes in the emission (and thus variability of the fluorescent line correlated with variability of the observed stellar plasma spectrum) or changes in the physical orientation of the emission (variability of the observed stellar plasma spectrum with \textit{no} change in the fluorescent line.

\section{Conclusions}
\label{sec:conclusion}

In this paper, we present and analyze new \textit{Chandra} gratings spectroscopy, new CCD-resolution \textit{XMM-Newton}/EPIC spectra, and archival CCD-resolution spectra from both \textit{XMM-Newton}/EPIC and \textit{Chandra}, of the benchmark Class I YSO HL Tau. Our results and conclusions are as follows:

\begin{itemize}
    \item We find that the HL Tau spectrum is consistently hot at all epochs, atypical for coronal sources in general but more common for typically-heavily-absorbed Class I YSOs.
    \item We find recurrent variability on a $\sim$21-day timescale in new observations of HL Tau with \textit{XMM-Newton} obtained in 2020. Based on forward modeling of the observations, the variability seen in 2020 can primarily be explained by the X-ray emission changing over time, rather than changing absorption of a stable X-ray spectrum over time. However, this is not necessarily the case at earlier epochs.
    \item It is unclear from our observations if the variability is stochastic in nature or not. Stochastic variability that manifests as it appears in the 2020 \textit{XMM-Newton} data would take the form of long, slow-rise, top-flat flaring. If the variability is in fact periodic, the most likely cause is an interaction between the star and disk, anchored to an orbiting feature in the disk inducing emission from the stellar surface.
    \item Emission lines in the \textit{Chandra}/HETG spectrum indicate the presence of both cool and hot plasma, with the pseudo-continuum of cool plasma hidden behind the high absorption.
    \item We detect faint emission from the cold Fe fluorescence line at 6.4 keV in the \textit{Chandra}/HETG spectrum. This is indicative of hot emission from the star stimulating fluorescence in cold neutral iron in the disk.
\end{itemize}

Additional observations are necessary to characterize the X-ray variability of HL Tau and, in particular, to further explore the potential for long-term recurrent variability. The Fe 6.4 keV line, which is produced by high-energy photons inducing fluorescence in cold iron in the disk, could be useful for differentiating between a change in the geometry of the emitting plasma versus a change in the intrinsic physical characteristics of the plasma.

\begin{acknowledgments}
We thank the anonymous referee for insightful comments that were invaluable to the science presented in this work.
S.M.S., S.J.W., and D.A.P. acknowledge support from the National Aeronautics and Space Administration through Chandra Award Number SAO GO8-19013X issued by the Chandra X-ray Observatory Center, which is operated by the Smithsonian Astrophysical Observatory for and on behalf of the National Aeronautics Space Administration under contract NAS8-03060. HMG acknowledges support from NASA grant 80NSSC21K0849. PCS acknowledges support from DLR 50OR2102.

This research has made use of data obtained from the Chandra Data Archive and the Chandra Source Catalog, and software provided by the Chandra X-ray Center (CXC) in the application packages CIAO and Sherpa.

Based on observations obtained with XMM-Newton, an ESA science mission with instruments and contributions directly funded by ESA Member States and NASA.

We acknowledge with thanks the variable star observations from the AAVSO International Database contributed by observers worldwide and used in this research.
\end{acknowledgments}

\vspace{5mm}
\facilities{CXO, XMM-Newton, AAVSO, ASAS-SN}

\software{astropy \citep{2013A&A...558A..33A,2018AJ....156..123A}, NumPy \citep{van2011numpy,harris2020array}, 
SciPy \citep{jones_scipy_2001,2020SciPy-NMeth}, 
Matplotlib \citep{Hunter:2007}, 
CIAO \citep{2006SPIE.6270E..1VF}, 
Sherpa \citep{2001SPIE.4477...76F}
}



\bibliography{references}{}
\bibliographystyle{aasjournal}



\end{document}